\documentclass[aps,prl,twocolumn,superscriptaddress,showpacs,floatfix]{revtex4}
\usepackage{graphicx,epsfig,amsmath}

\newcommand{\bohrm}{\mbox{$\mu_{B}$}}

\newcommand{\etal}{\textit{et al.}}

\newcommand{\AlxGaAs}[2]{\mbox{$\text{Al}_{#1}\text{Ga}_{#2}\text{As}$}}

\newcommand{\ZeemanE}{\mbox{$|g|\bohrm B$}}

\newcommand{\startsubfig}[2]{Figure~\ref{fig:#1}(#2)}
\newcommand{\subfig}[2]{Fig.~\ref{fig:#1}(#2)}
\newcommand{\allfig}[1]{Fig.~\ref{fig:#1}}

\begin{document}


\title{Spin-Dependent Tunneling of Single Electrons into an Empty Quantum Dot}

\author{S. Amasha}
	\email{samasha@mit.edu}
	\affiliation{Department of Physics, Massachusetts Institute of Technology, Cambridge, Massachusetts 02139}

\author{K. MacLean}
	\affiliation{Department of Physics, Massachusetts Institute of Technology, Cambridge, Massachusetts 02139}

\author{Iuliana P. Radu}
	\affiliation{Department of Physics, Massachusetts Institute of Technology, Cambridge, Massachusetts 02139}

\author{D. M. Zumb\"{u}hl }
	\affiliation{Department of Physics and Astronomy, University of Basel, Klingelbergstrasse 82, CH-4056 Basel, Switzerland}

\author{M. A. Kastner} 
	\affiliation{Department of Physics, Massachusetts Institute of Technology, Cambridge, Massachusetts 02139}
	
\author{M. P. Hanson} 
	\affiliation{Materials Department, University of California, Santa Barbara 93106-5050}	

\author{A. C. Gossard} 
	\affiliation{Materials Department, University of California, Santa Barbara 93106-5050}


\begin{abstract}
	Using real-time charge sensing and gate pulsing techniques we measure the ratio of the rates for tunneling into the excited and ground spin states of a single-electron AlGaAs/GaAs quantum dot in a parallel magnetic field. We find that the ratio decreases with increasing magnetic field until tunneling into the excited spin state is completely suppressed. However, we find that by adjusting the voltages on the surface gates to change the orbital configuration of the dot we can restore tunneling into the excited spin state and that the ratio reaches a maximum when the dot is symmetric.
\end{abstract}


\maketitle

	
	The spin physics of laterally gated quantum dots in AlGaAs/GaAs heterostructures is of great interest \cite{Hanson2006:SpinsinQD} because of potential uses for quantum dots in spin-based applications. A laterally gated quantum dot consists of electrons in a two-dimensional electron gas (2DEG) at the AlGaAs/GaAs interface that have been confined in a potential defined by surface gates. The ability to manipulate \cite{Petta2005:CoherentManipulation, Koppens2006:ESR} and read out \cite{Elzerman2004:SingleShotReadOut, Hanson2005:SpinDepTunnelRates} the spins of these confined electrons has made this type of quantum dot promising for applications in quantum information processing \cite{Loss1998:QDotQuanComp}. Electron tunneling between the quantum dot and the remaining 2DEG regions has been studied at zero magnetic field \cite{Gustavsson2006:CountStat, MacLean2007:WKB}, as well as in fields perpendicular \cite{Ciorga2000:AddSpec, Hitachi2006:SpinSelSpec, Zarchin2007:ElecBunching} and parallel \cite{Lindemann2002:SpinStability} to the 2DEG, and dots have been developed as spin filters \cite{Folk2003:BidirSF, Hanson2004:BipolarSF} and a spin pump \cite{Watson2003:SpinPump} for applications in spintronics \cite{Wolf2001:Spintronics, Zutic2004:Spintronics, Awschalom2007:spintronics}. Tunneling into self-assembled InAs quantum dots coupled to three-dimensional electron reservoirs has also been studied in magnetic fields \cite{HapkeWurst2000:BsingTunnel, Vdovin2007:MagnetoAniso}. Despite this progress in understanding the spin physics of dots, measurements of the spin states of electrons emitted from a lateral quantum dot in the Coulomb blockade regime by Potok \etal~\cite{Potok2003:PolCurrent} remain unexplained. Using a magnetic focusing geometry and a quantum point contact spin sensor, Potok \etal~have measured the spin polarization of electrons emitted from a quantum dot as the dot's spin state is varied from $S= 0$ to $S= 1$. Surprisingly, these authors find no variation in the polarization of the emitted electron spin as they varied the spin state of the dot. These experiments point out the need to further probe the spin-dependence of tunneling in quantum dots.
	
	In this Letter we use real-time charge sensing and gate pulsing techniques to measure tunneling into the spin states of a single-electron quantum dot whose levels have been split by a parallel magnetic field. We find that the ratio of the rates for tunneling into the excited and ground spin states of the empty dot depends on the magnetic field and the orbital configuration. Specifically, we find that the ratio decreases with increasing magnetic field until tunneling into the excited spin state is completely suppressed. However, we find that by adjusting the voltages on the surface gates to change the shape of the dot, we can restore tunneling into the excited spin state and that the ratio of the tunneling rates reaches a maximum when the dot is symmetric.
	
\begin{figure}
\begin{center}
\includegraphics[width=8.0cm, keepaspectratio=true]{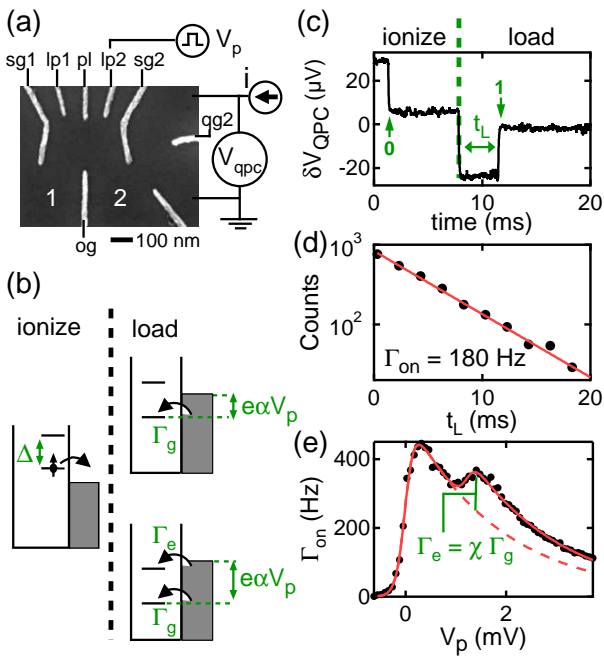}
\end{center}

\caption{(color online) (a) Electron micrograph of the gate geometry. Negative voltages are applied to the labeled gates while the unlabeled gate and the ohmic leads are kept at ground. Voltage pulses are applied to gate LP2. (b) Dot energy diagrams showing the position of the Zeeman states during the pulse sequence. (c) Example of real-time data. The direct capacitive coupling between LP2 and the QPC causes the QPC to respond to the pulse sequence; electron tunneling events are evident on top of this response. The 0 denotes when an electron tunnels off the dot, while the 1 denotes when an electron tunnels on. (d) Example of a histogram of $t_{L}$ for a given pulse depth. Fitting these data to an exponential (solid line) gives $\Gamma_{on}$. (e) $\Gamma_{on}$ as a function of pulse depth $V_{p}$ at $B= 5$ T.  The solid and dashed lines are fits discussed in the text to obtain $\chi= \Gamma_{e}/\Gamma_{g}$.
}
\label{fig:fig1}
\end{figure}

	We fabricate our dots from an \AlxGaAs{0.3}{0.7}/GaAs heterostructure grown by molecular beam epitaxy. The two-dimensional electron gas (2DEG) formed at the material interface $110$ nm below the surface has a density of $2.2\times10^{11}$ $\text{cm}^{-2}$ and a mobility of $6.4\times10^{5}$ $\text{cm}^{2}$/Vs \cite{Granger2005:TwoStageKondo}. We pattern metallic gates on the surface of this heterostructure as shown in \subfig{fig1}{a}. By applying negative voltages to the labeled gates we deplete the 2DEG underneath them and form a single dot containing one electron, as well as a quantum point contact (QPC) between gates SG2 and QG2. The remaining 2DEG regions form the ohmic leads, two of which are numbered in \subfig{fig1}{a}. Electrons tunnel onto and off of the dot through the tunnel barrier defined by gates SG2 and OG while the tunneling rate through the SG1-OG barrier is kept negligibly small. To measure the occupancy of the dot, we use the QPC as a charge sensor \cite{Field1993:NoninvasiveProbe}. When an electron tunnels onto or off of the dot, it changes the resistance of the QPC and we detect this change by sourcing a current $I$ and measuring the change in voltage $\delta V_{QPC}$. By making the tunneling rate slower than the bandwidth of our circuit, we observe electron tunneling events in real time \cite{Elzerman2004:SingleShotReadOut, Lu2003:RealTimeDet, Gustavsson2006:CountStat, MacLean2007:WKB}. All measurements have been made in a dilution refrigerator with an electron temperature $T= 120$ mK. The magnetic field $B$ is applied parallel to the 2DEG and splits the spin states of the dot by an energy $\Delta= \ZeemanE$, where $|g|= 0.39$ and is discussed below.

	To measure $\Gamma_{on}$, which is the rate at which electrons tunnel onto the dot, we use the two step pulse sequence shown in \subfig{fig1}{b}. The first step is to ionize the dot by pulsing gate LP2 to bring both spin states above the Fermi energy of the lead, so that any electron on the dot tunnels off. During the subsequent load step we apply a positive pulse voltage $V_{p}$ to bring the ground Zeeman state below the Fermi energy of the lead by an energy $E_{p}= e\alpha V_{p}$, where $e\alpha$ is a conversion constant we have calibrated separately \cite{Amasha2007:SizeDepT1, Gustavsson2006:superPoisson}. If only the ground spin state is below the Fermi energy (top diagram in \subfig{fig1}{b}) then $\Gamma_{on}$ is equal to the rate for tunneling into the ground state $\Gamma_{g}$. For large enough $V_{p}$, the excited spin state is also below the Fermi energy (bottom diagram in \subfig{fig1}{b}), and then $\Gamma_{on}= \Gamma_{g} + \Gamma_{e}$, where $\Gamma_{e}$ is the rate for tunneling into the excited spin state.

	\startsubfig{fig1}{c} shows an example of a pulse sequence taken with our real-time charge detection system. During the ionization step, an electron tunnels off the dot and then during the load step, an electron tunnels back onto the dot. For each pulse sequence, we measure the time $t_{L}$ between when the dot is pulsed into the load state and when an electron tunnels onto the dot. \startsubfig{fig1}{d} shows a histogram of these data for a fixed $V_{p}$. By fitting these data to an exponential we extract $\Gamma_{on}$ at this value of $V_{p}$. \startsubfig{fig1}{e} shows an example of $\Gamma_{on}$ as a function of $V_{p}$. The large increase at $V_{p}= 0$ corresponds to the ground Zeeman state passing the Fermi level, while the increase at $V_{p}\approx 1.5$ mV corresponds to the excited Zeeman state passing the Fermi level.  

  In MacLean \etal~\cite{MacLean2007:WKB} we showed that the tunnel rate into an empty dot state at $B= 0$ can be described by $\Gamma= \Gamma_{0} e^{-\beta V_{p}} f(-e\alpha V_{p})$. Here $\Gamma_{0}$ is the tunnel rate through the SG2-OG tunnel barrier when the energy of the dot state is aligned with the Fermi energy of the lead, and the exponential factor describes the decrease in the tunnel rate as $V_{p}$ pulls the energy of the dot level further below the top of the tunnel barrier. The Fermi function $f(E)= (1+e^{E/k_{B} T})^{-1}$ describes the occupation of states in the lead at the energy of the dot state. In a magnetic field the spin states of the dot are split by $\Delta$ and we can describe $\Gamma_{on}$ by
\begin{equation}
  \Gamma_{on}= \Gamma_{0} e^{-\beta V_{p}} [f(-e\alpha V_{p})+ \chi f(-e\alpha V_{p}+\Delta)]
\label{eqn:GamEq}	
\end{equation}
where $f(-e\alpha V_{p})$ and $f(-e\alpha V_{p}+\Delta)$ describe the occupation of the lead at the energies of the ground and excited spin states, respectively, and the factor $\chi$ accounts for spin-dependent tunneling. The solid line in \subfig{fig1}{e} shows a fit to Eqn. \ref{eqn:GamEq}, and there is good agreement with the data. The dashed line shows the contribution of tunneling into the ground state, given by $\Gamma_{g}= \Gamma_{0} e^{-\beta V_{p}} f(-e\alpha V_{p})$, while the remaining contribution is caused by tunneling into the excited state $\Gamma_{e}$. When both spin states are below the Fermi energy such that $f(-e\alpha V_{p}) \approx f(-e\alpha V_{p}+\Delta) \approx 1$ then $\chi= \Gamma_{e}/\Gamma_{g}$. Hence from the fit, we can extract $\chi$.

\begin{figure}
\begin{center}
\includegraphics[width=8.0cm, keepaspectratio=true]{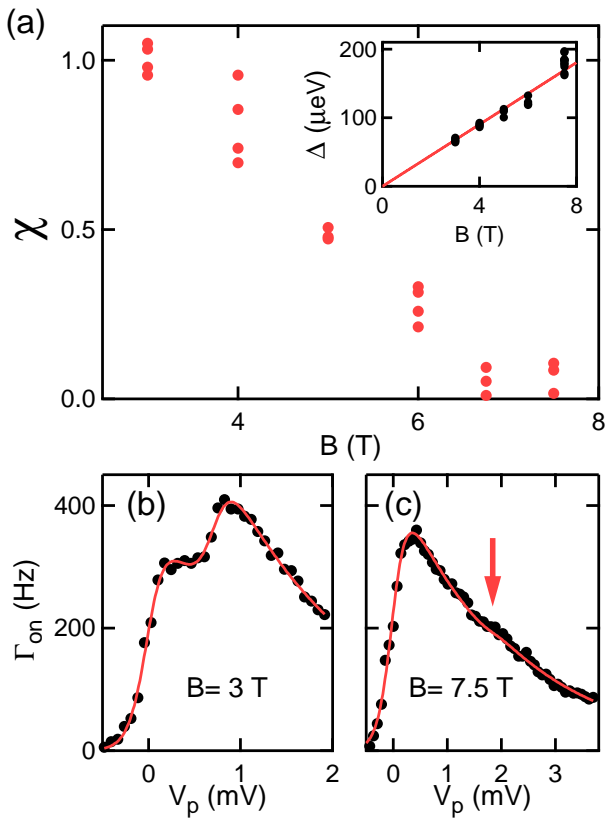}
\end{center}

\caption{(color online) (a) $\chi$ as a function of magnetic field from fits to data such as those in \subfig{fig1}{e}. For $B\leq 6$ T, the excited state feature is clearly visible and $\Delta$ can be extracted from the fit. For $B> 6$ T, the feature is not visible and fits are performed fixing $\Delta= \ZeemanE$, where $|g|= 0.39$ is determined by fitting measurements from which we can extract $\Delta$ (inset). These measurements include values at $B= 7.5$ T for different orbital configurations where tunneling into the excited spin state is not suppressed. (b) Examples of data at $B= 3$ T. The increase in tunnel rate caused by the excited state passing below the Fermi energy is clearly visible. (c) Example of data at $B= 7.5$ T. The arrow marks the value of $V_{p}= \Delta/e\alpha$ where the feature is expected to be.
}
\label{fig:fig2}
\end{figure}

	\startsubfig{fig2}{a} shows measurements of $\chi$ obtained by fitting lineshapes at different applied magnetic fields. \startsubfig{fig2}{b} and (c) show examples of data at $B= 3$ and $7.5$ T, respectively. In \subfig{fig2}{b} the increase in $\Gamma_{on}$ caused by the excited spin state passing the Fermi energy is clearly visible. In \subfig{fig2}{c} no increase is visible; the arrow marks the value of $V_{p}= \Delta/e\alpha$ where the feature should be. From \subfig{fig2}{a}, we see that application of a magnetic field suppresses the tunneling rate $\Gamma_{e}$ into the excited spin state, relative to that into the ground state \cite{footnote:Rates}.

\begin{figure}
\begin{center}
\includegraphics[width=8.0cm, keepaspectratio=true]{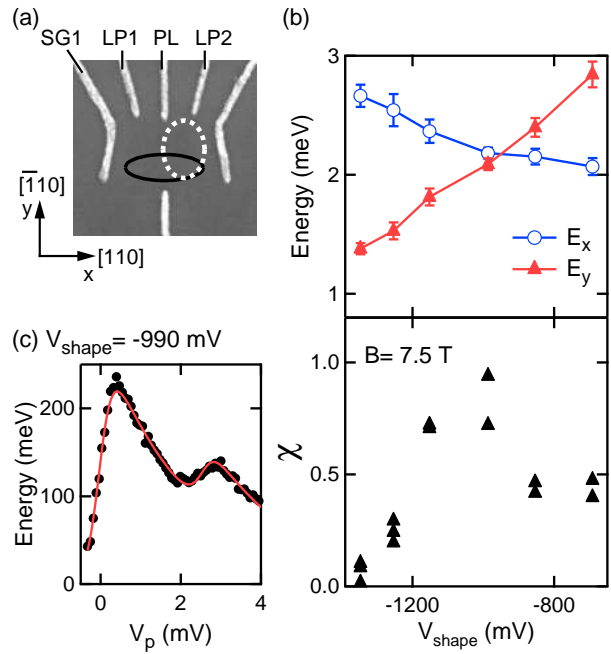}
\end{center}

	\caption{(color online) (a) The black solid (white dotted) ellipse illustrates the expected dot shape for less (more) negative $V_{shape}$. For all data in this paper, the magnetic field is applied along the $y$-axis. (b) The top panel shows the energy spectrum of the excited orbital states as a function of $V_{shape}$. The bottom panel shows $\chi$ measured at $B= 7.5$ T for each value of $V_{shape}$. (c) Data at $V_{shape}= -987$ mV and $B= 7.5$ T. Unlike \subfig{fig2}{c}, at this value of $V_{shape}$ the excited state feature is clearly present. The value of $V_{p}$ at which it appears is different than in \subfig{fig2}{c} because the conversion constant $e\alpha$ changes with $V_{shape}$. }
\label{fig:fig3}
\end{figure}

	We can change this suppression by varying the voltages on the gates that define the dot \cite{Amasha2007:SizeDepT1}. \startsubfig{fig3}{a} defines the $x$ and $y$ axes, which are aligned with the $[110]$ and $[\overline{1}10]$ GaAs crystalline axes, respectively. When the voltages on all dot gates are approximately equal, we expect from the gate geometry that the dot is less confined along $x$ than along $y$ (black solid ellipse in \subfig{fig3}{a}). We change the shape of the dot by applying a more negative voltage to SG1, which pushes the dot wavefunction toward SG2, thereby increasing confinement along $x$. Simultaneously, we make the voltages on LP1, PL, and LP2 less negative which reduces confinement along $y$ (white dotted ellipse in \subfig{fig3}{a}), while keeping the ground state energy constant. We parameterize a set of gate voltages by $V_{shape}$, the numeric value of which is the voltage on gate SG1.
	
	To characterize the change in shape of the dot we use the energy of the excited orbital states \cite{Amasha2007:SizeDepT1}. We can model the electrostatic potential of the dot with an anisotropic harmonic oscillator potential $U(x,y) = \frac{1}{2} m^{*} \omega_{x}^{2} x^{2} + \frac{1}{2} m^{*} \omega_{y}^{2} y^{2}$. Then the energies of the excited orbital states relative to the ground orbital state are determined by the confinement: $E_{x}= \hbar\omega_{x}$ and $E_{y}= \hbar\omega_{y}$. For less negative $V_{shape}$ the dot is less confined along $x$ than along $y$ so we expect $E_{x}<E_{y}$. As we make $V_{shape}$ more negative we increase the confinement along $x$ and decrease the confinement along $y$, and so we expect that $E_{x}$ should increase and $E_{y}$ should decrease as $V_{shape}$ is made more negative. 
	
	The top panel of \subfig{fig3}{b} shows the energies of the first two excited states of the dot measured using gate pulsing and real-time charge detection techniques \cite{Amasha2007:SizeDepT1}. As expected, the energy of one state increases and that of the other state decreases as $V_{shape}$ is made more negative, and this allows us to identity the states as indicated in the top panel of \subfig{fig3}{b}. At each value of $V_{shape}$ we perform measurements of $\chi$ at $B= 7.5$ T and the results are shown in the bottom panel of \subfig{fig3}{b}. We also extract $\Delta$, and have verified that it is independent of $V_{shape}$. The data in \allfig{fig2} have been taken at the most negative value of $V_{shape}= -1350$ mV. Making $V_{shape}$ less negative (\subfig{fig3}{b}) changes $\chi$, and $\chi$ reaches a maximum of $\approx 1$ at $V_{shape} \approx -990$ mV when the dot is symmetric. \startsubfig{fig3}{c} shows data taken at this $V_{shape}$. In contrast to \subfig{fig2}{c} the excited state is now clearly visible.

	If we assume that tunneling is elastic \cite{MacLean2007:WKB} and that there is no coupling between the electron orbital and spin states in the dot or the leads, we would expect that $\chi = 1$. This is because in the absence of such coupling the excited and ground spin states of the dot have the same orbital wavefunction and hence the same overlap with the leads. That we observe $\chi$ changing with the magnetic field and with the shape of the dot implies that this simple picture does not adequately describe the physics of electron tunneling in a magnetic field. 
	
	We have considered several possible mechanisms- a perpendicular magnetic field, the spin-orbit interaction, and interaction with the QPC- but have found that none of these account for the observed spin-dependence of tunneling. Although we orient the sample such that the field is parallel to the 2DEG, a small misalignment could give a perpendicular field $B_{\bot}$. We estimate that the sample is parallel to within $5$ degrees and this limits $B_{\bot} < 0.65$ T at $B= 7.5$ T, which is the highest field we use. Since we are measuring single-electron tunneling into an empty quantum dot, there are no exchange effects in the dot; rather, the dot states are single-particle states. But $B_{\bot}$ can affect the states in the ohmic leads by forming Landau levels, and one possibility is that we would observe spin-dependent tunneling were the dot a spin-sensitive probe of the states in the leads \cite{Ciorga2000:AddSpec}. We do not believe this is the case for several reasons. First, this mechanism does not explain how changes in the dot shape could affect $\chi$. Also, we have observed spin-dependent tunneling in a second device where we have checked for $B_{\bot}$ by sourcing
a current through the 2DEG and measuring the Hall voltage. Based on this measurement, we estimate $B_{\bot} \approx 20$ mT at $B= 7.5$ T.
	
	We have also considered the effects of the spin-orbit interaction (SOI) in both the dot and the leads. The effect of the SOI on the dot states is small because it is on the order of $x/\lambda_{SO}\approx 8\times 10^{-3}$ where $x\approx 17$ nm is the length scale for a harmonic oscillator potential approximating a dot with energy spacing $E \approx 2$ meV, and the spin-orbit length $\lambda_{SO}\sim 2~\mu$m describes the strength of the SOI \cite{Amasha2007:SizeDepT1, Zumbuhl2002:SOCoupling}. In the leads the SOI can be thought of as a momentum-dependent effective magnetic field $B_{SO}$ which is $\approx 6$ T at the Fermi energy using $\lambda_{SO}\sim 2~\mu$m. As the magnetic field increases we expect the Zeeman splitting to begin to dominate the SOI and the physics would approach the simple picture discussed above, so that $\chi$ should approach $1$ at high fields. This is not what we observe. 
	
	Finally, we have checked whether the spin-dependence of tunneling depends on the current in the QPC by measuring $\chi$ for several different currents through the QPC in a second device. We have varied the current by a factor of $3$ (from $0.9$ to $2.7$ nA) but observed no significant variation in $\chi$. These observations suggest that the QPC is not responsible for the observed effect.
	
	In summary, we measure the ratio of the rates for a single electron tunneling into the excited and ground spin states of an empty quantum dot, which is a particularly simple situation where exchange and correlation effects that can occur in multi-electron dots are not present. Surprisingly, we find that the ratio decreases with increasing magnetic field and that the ratio reaches a maximum when the dot is symmetric. We know of no theoretical explanation for these observations, which underscores the fact that understanding the spin-dependence of tunneling continues to be an important open problem in the physics of quantum dots.

	We are grateful to D. Loss, L. Levitov, E. I. Rashba, and J. B. Miller for discussions and to I. J. Gelfand and T. Mentzel for experimental help. This work was supported by the US Army Research Office under W911NF-05-1-0062, by the National Science Foundation under DMR-0353209, and in part by the NSEC Program of the National Science Foundation under PHY-0117795.
	

\begin{thebibliography}{28}
\bibitem[{foo()}]{footnote:Rates}
\bibinfo{note}{The bare tunneling rate onto the dot $\Gamma_{0}$ changes with
  magnetic field and hence it is necessary to adjust this rate using the gate
  voltages to keep $\Gamma_{on}$ at measurable levels ($\approx 200-400$ Hz at
  the peak). This means it is not possible to compare the values of
  $\Gamma_{on}$ at two different magnetic fields. Consequently we cannot
  determine whether the decrease in $\chi$ is caused by an increase in
  $\Gamma_{g}$ or a decrease in $\Gamma_{e}$.}
\end{thebibliography}

\end{document}